\documentclass[letterpaper, 10 pt, conference]{ieeeconf}  
\IEEEoverridecommandlockouts
\usepackage{cite}
\usepackage{amsmath,amssymb,amsfonts}
\usepackage{bm} 
\usepackage{algorithm}
\usepackage{algorithmic}
\usepackage{graphicx}
\usepackage{textcomp}
\usepackage{xcolor}
\usepackage{todonotes}
\def\BibTeX{{\rm B\kern-.05em{\sc i\kern-.025em b}\kern-.08em
    T\kern-.1667em\lower.7ex\hbox{E}\kern-.125emX}}
    
\usepackage{booktabs}
\usepackage{cleveref}
\crefformat{equation}{(#2#1#3)}
\crefrangeformat{equation}{(#3#1#4) to~(#5#2#6)}
\Crefname{figure}{Fig.}{Figs.}
\Crefname{tabular}{Tab.}{Tabs.}
\usepackage{siunitx}

\newenvironment{bsmallmatrix}
  {\left[\begin{smallmatrix}}
  {\end{smallmatrix}\right]}

\begin{document}

\title{\LARGE \bf The Syncline Model - Analyzing the Impact of Time Synchronization in Sensor Fusion*}

\author{Erling R. Jellum$^{1\dagger}$, Torleiv H. Bryne$^{1}$, Tor Arne Johansen$^{1}$ and Milica Orlandić$^{2}$
\thanks{* This work was supported by the Research Council of Norway (RCN) through the projects Machine Piloted Unmanned Systems (MPUS), grant number 309328, Multi-Sensor Data Timing, Synchronization and Fusion for Intelligent Robots, grant number 327538, and the center of excellence NTNU AMOS, grant number 223254.}
\thanks{$^{1}$Norwegian Univ. Sci\&Tech, Center for Autonomous Marine Operations and Systems, Dept. Engineering Cybernetics} 
\thanks{$^{2}$Norwegian Univ. Sci\&Tech, Dept. Electronic Systems}
\thanks{$\dagger$Email corresponding author: erling.r.jellum@ntnu.no.}
}

\maketitle
\begin{abstract}
The accuracy of sensor fusion algorithms are limited by either the intrinsic sensor noise, or by the quality of time synchronization of the sensors.
While the intrinsic sensor noise only depends on the respective sensors, the error induced by quality of, or lack of, synchronization depends on the dynamics of the vehicles and robotic system and the magnitude of time synchronization errors. To meet their sensor fusion requirements, system designers must consider both which sensor to use and also how to synchronize them.
This paper presents the Syncline model, a simple visual model of how time synchronization affects the accuracy of sensor fusion for different mobile robot platform.
The model can serve as a simple tool to determine which synchronization mechanisms should be used.
\end{abstract}

\section{Introduction}
Sensor fusion algorithms are often developed with, and evaluated on, data sets with perfectly synchronized sensor measurements, e.g. the EuRoC data set~\cite{euroc}.
In reality, separate sensors are never completely synchronized. 
The degree to which they are synchronized depends on the which synchronization primitives are used and how they are used.
There exists few commercial products for flexible synchronizing various sensor sources.
System designers are typically left to implement their own synchronization solutions \cite{micromob}.
Not only is this error-prone, but they lack a framework for understanding the precision requirements for such a solution.
For extreme precision, one could implement a custom timestamping circuit on an FPGA yielding deterministic timestamping and parsing of sensor data.
In the other end of the scale, there is software timestamping of sensor samples ``on arrival'' running under some non-real time operating system.
The first approach can achieve synchronization precision in the order of nanoseconds, while the second solution can suffer synchronization precision in the 100s of milliseconds range~\cite{gnss_latency_hansen}.

A passive algorithm for estimating synchronization error is proposed in~\cite{passive_sync} and algorithms for incorporating precision- and jitter-estimates in the sensor fusion algorithms in~\cite{ins_sync}. 

The accuracy of the sensor fusion algorithm is affected by both the intrinsic sensor noise and the accuracy of the timestamping of the sensor measurements, i.e. the synchronization precision.
The degree of which the synchronization precision affects the overall accuracy of the sensor fusion algorithm is dependent on the dynamics of the system.
For a system with slow dynamics, e.g. an unmanned surface Vehicle (USV), reducing synchronization precision from nanoseconds to microseconds does affect the overall accuracy significantly. I.e. the intrinsic sensor noise is the limiting factor.
However, for a system with fast dynamics, e.g. a unmanned aerial vehicles (UAVs), using more accurate sensors does not improve the sensor fusion accuracy significantly if the synchronization precision is only in the order of hundreds of milliseconds. I.e. the synchronization precision is the limiting factor.

In this work we introduce the Syncline model which quantifies the effects of synchronization precision and instrinsic sensor noise and provides a clear and insightful visual performance model for system designers integrating sensors on mobile robots. The model is applied on two sensor fusion applications using realistic sensor and robot platform models:
\begin{enumerate}
    \item Direct georeferencing from either UAV, USV, autonomous underwater vehicle (AUV), or surface vessels, and
    \item Subsea survey with a surface vessel, and an AUV.
\end{enumerate}
Analytical models are derived for both examples and compared to the predictions of the Syncline model.

This paper is organized as follows. 
In Section II we give a brief background into sensor synchronization.  
Section II introduced related work. 
Section IV introduces the mathematical notation and coordinate frames used.
Section V derives the general Syncline model mathematically before it is applied on two concrete examples in Section VI. 
The work is concluded in Section VII.

\section{Background}

\subsection{Sensor synchronization primitives}
Commercial off-the-shelf sensors typically include analog-to-digital conversion and have a digital communication interface to the host.
The synchronization primitives exposed by these sensors typically fall in one of the following categories:
\begin{enumerate}
    \item [\textbf{1.}] \textbf{1PPS}~\cite{1PPS}
     These sensors typically include an embedded microcontroller which samples and timestamps sensor measurements based on its own internal clock. A 1PPS is a 1-wire input signal which expects a flank at the beginning of each second. The embedded microcontroller can then timestamp measurements relative the beginning of a whole second. Some sensors  can receive and parse certain global navigation satellite systems (GNSS) timing packets from a GNSS receiver by the use of a UART receiver. This allows the sensor to timestamp measurements on the Coordinated Universal Time (UTC) timeline.
    \item [\textbf{2.}]\textbf{Network synchronization}. High-end sensors will typically include microcontrollers or microprocessors. They can run the TCP/IP stack and support IEE1588 Precision Time Protocol (PTP) \cite{iee1588} or Network Time Protocol (NTP) \cite{NTP}  which synchronizes the sensors' clock relative a grandmaster clock on the network.
     \item [\textbf{3.}]\textbf{Time-of-validity (TOV) synchronization}. 
       Some sensors do not timestamp their measurements, but rather expose timing information to the host.
       A TOV signal is a flank on a single wire outputted by the sensor at the precise moment of sampling.
       The host must detect and timestamp the TOV signal and associate it with the proceeding measurement.
     \item [\textbf{4.}]\textbf{Triggered synchronization}. Other sensor, like cameras and some inertial measurement units (IMU), can be triggered to provide a measurement sample. This is normally done through a single wire connected to the sensor. The user can generate a flank on this pin and associate the time of the flank with the proceeding measurement.
\end{enumerate}
\subsection{Timestamping methods}
There are multiple ways for a computational device to use these synchronization primitives.

\begin{enumerate}
\item [\textbf{1.}]\textbf{Hardware timestamping}
To achieve deterministic timestamping accuracy the only solution is to perform it in hardware. This applies to 1PPS, TOV and Triggered synchronization. This means using dedicated Timer peripherals to timestamp and generate events. The accuracy of hardware timestamping is typically limited by the frequency of its clock source and that clock's drift.
\item [\textbf{3.}]\textbf{GPIO-based timestamping}
In GPIO-based timestamping interrupts are used to detect and timestamp input events (TOV).
To generate output events (1PPS and Triggers) one can either configure a timer interrupt or just use the sleep function if one uses a multi-threaded platform.
GPIO-based timestamping is only deterministic under strong restriction.
Unless the timestamping interrupt routine has the highest priority and interrupts never are disabled, like during system-calls in Linux, GPIO-based timestamping will not give deterministic timestamping.
\item [\textbf{3.}]\textbf{Timestamp-on-arrival}
Another option is to just timestamp the sensor measurements as they arrive to the compute platform. This might be the only choice if either your sensor has no synchronization primitives, or your platform has no GPIOs. Sometimes UART to USB adapters are used to easily integrate sensors using UART/RS232/RS422 with a compute platform over USB. The ubiquitous FTDI UART-USB chips are known to introduce up to 16ms buffering delay \cite{ftdi_latency}.
\end{enumerate}

\section{Preliminaries}
\subsection{Mathematical Notations}

The Euclidean vector norm is denoted \(\|\cdot\|_2\), and the \(n \times n\) identity matrix is denoted \(\bm{I}_n\). The transpose of a vector or a matrix is denoted \((\cdot)^\intercal\). Coordinate frames are expressed as \(\{\cdot\}\), while \(\bm z^a_{bc} \in \mathbb{R}^3\) denotes a vector \(\bm z\) from frame \(\{b\}\) to \(\{c\}\), resolved in \(\{a\}\). 
\(\bm S(\cdot) \in SS(3)\) denotes a skew symmetric matrix such that $\bm S(\cdot) = -\bm S^\intercal(\cdot)$ and $\bm S(\bm z_1)\bm z_2 = \bm z_1 \times \bm z_2$ and $\bm z_1 \cdot \bm z_2$ is a dot product for the two vectors \(\bm z_1,\bm z_2 \in \mathbb{R}^3\).

\subsection{Attitude representations and relationships}
Table \ref{tab:frames} introduces the coordinate frames used in this paper. 
The rotation matrix, \(\bm R_{ab} \in SO(3)\), represents the rotation between $\{a\}$ and $\{b\}$ frames such that $ \bm z^a_{\star \star} = \bm R_{ab}\bm z^b_{\star \star}  $.

Additionally, the Euler angles roll, pitch and yaw are defined as $\phi,\theta,\psi$
and relate to rotation matrix from body to NED using:
\begin{equation}
    \bm R_{nb}(\bm \Theta) = 
        \begin{bmatrix}
        c\theta c\psi & -c\phi s\psi + s\phi s\theta c\psi &  s\phi s\psi + c\phi s\theta c\psi \\
        c\theta s\psi &  c\phi c\psi + s\phi s\theta s\psi & -s\phi c\psi +  c\phi s\theta s\psi \\
        -s\theta      & s\phi c\theta                      &  c\phi c\theta
        \end{bmatrix}
        \label{eq:R_nb}
\end{equation}
where \(c\star\) denotes \(\cos(\star)\) and \(s\star\) denotes \(\sin(\star)\) and
\begin{equation}
    \bm \Theta_{nb} = 
    \begin{bmatrix}
        \phi & \theta &\psi
    \end{bmatrix}^\top.
\end{equation}
For an estimated or measured rotation matrix $\hat{\bm R}_{nb}$, the true rotation matrix $\bm R_{nb}$ relates to the attitude error $\bm \epsilon_a$
\begin{equation}
    \bm R_{nb} \approx \hat{\bm R}_{nb}\left( \bm I_3 + \bm S(\bm \epsilon_a)\right)
\end{equation}
if  $\bm \epsilon_a$ is small.
Hence,
\begin{align}
    \bm R_{nb}\left( \bm I_3 + \bm S(\bm \epsilon_a)\right)^{-1} &\approx \hat{\bm R}_{nb} \notag \\
    \Rightarrow \hat{\bm R}_{nb} &\approx  \bm R_{nb}\left( \bm I_3 - \bm S(\bm \epsilon_a)\right)
\end{align}
due to $\left( \bm I_3 + \bm S(\bm \epsilon_a) \right) ^{-1} \approx \left( \bm I_3 - \bm S(\bm \epsilon_a) \right)$ for small $\bm \epsilon_a.$
\begin{table}[tb]
    \centering
    \caption{\label{tab:frames}Coordinate frames}
    \vspace*{-0.2cm}
    \begin{tabular}{ ll }
        \toprule
        Symbol & Name \\
        \midrule 
       $e$ & Earth Centered Earth Fixed (ECEF)\\
       $n$ & North East Down (NED)\\
         $b_{r}$ & Mobile robot reference frame\\
         $b_{o}$ & Object of interest\\
         $b_{l}$ & LiDAR\\
         $b_{g}$ & Global navigation satellite system (GNSS) receiver\\
         $b_{sv}$ & Surface Vessel reference frame\\
         $b_{uv}$ & Autonomous underwater vehicle (AUV) reference frame\\
         $b_{urx}$ & Hydro acoustic ultra short baseline(USBL) receiver\\
         $b_{utp}$ & USBL transponder\\
         $b_{mbe}$ & Multibeam echosounder (MBE) \\
         \bottomrule
    \end{tabular}
    \vspace*{-0.2cm}
\end{table}%
\section{Related Work}
State estimation subject to non-deterministic delays and clock synchronization errors have been extensively studied, see \cite{networked_state_est2}.

In \cite{roofline} the authors introduce the Roofline model, a simple visual performance model for computational platforms.
Their key insight is that the performance of a certain application running on a computational platform is either limited by its memory bandwidth or its raw compute capabilities.
The Syncline model builds on this idea.
Instead of being limited by either memory bandwidth or computing power, mobile robots are limited by either sensor accuracy or synchronization quality.

The F1 Roofline model \cite{f1_roofline} is an adaptation of the Roofline model for multi rotor UAVs. They model the max flights speed of a multi rotor as either limited by sensor accuracy, processing speed or robot weight.
They show that for some applications and robots using a slower and lighter compute platform could increase the max flight speed.

The Syncline model is inspired by this line of work.

\section{Syncline Model}
The Syncline model considers three kinds of sensor modalities: 

\begin{enumerate}
  \item Sensors measuring the 3D localization of the robot, e.g. GNSS,
  \item Sensors measuring the attitude or orientation of the robot, e.g. INS, and
  \item Sensors measuring range and bearing to other objects, e.g. LiDAR, camera, Radar, Sonar etc.
\end{enumerate}

Some of the modalities will actually have internal sensor fusion which we will not consider. 

\subsection{State-space formulation}
Consider a dynamic system represented by the following nonlinear state-space equation,
\begin{equation}
  \bm{\dot{x}}(t) = \bm f(\bm{x}(t), \bm{u}(t)),
\end{equation}
where $\bm x$ is the state vector and $\bm u$ is a vector with input signals.
This could represent any type of system which needs accurately timestamped measurements in order to perform sensor fusion.

The focus of this paper is on mobile robots, in that case $\bm{x}= (\bm{p}_{eb}^{e},\bm \Theta_{nb}, \dots)^\intercal$.
Here $\bm{p}_{eb}^{e}$ is the position of the robots \textit{body} with respect to Earth's center relative defined by the \textit{ECEF} coordinate frame.
$\bm{\Theta}_{nb}$ is the attitude of the robot which can be given via Euler angles. The rest of the state variables depend on the robotic system and its environment. 
\subsection{Sensor measurement model}
The sensor measurements are modelled as the system output as follows
\begin{align}
    \bm y(t_{k}) = \bm{h}(\bm{x}(t_{k}+\mu_\mathrm{sync})) + \bm \epsilon_\mathrm{sensor},
    \label{eq:state_space_y}
\end{align}
here $\bm y (t_{k})$ is a sensor measurement timestamped with the time tag $t_{k}$.
The measurement is assumed to be a function of the state vector $\bm h(\bm{x}(t_{k}))$ with additional sensor noise denoted $\bm \epsilon_\mathrm{sensor}$.
The independent random variable $\mu_\mathrm{sync}$ represents the \textit{synchronization error}.
The synchronization error leads to inaccurate timestamping as a sensor measurement valid at time $t_k+\mu_\mathrm{sync}$ is associated with timestamp $t_k$. 
The variables $\bm \epsilon_\mathrm{sensor}$ and $\mu_\mathrm{sync}$ are assumed to be uncorrelated variables.

\subsection{System dynamics}
The potential consequences of synchronization errors between sensor sources is related to the \textit{system dynamics} of the robot.
The actual error introduced depends on how much the measured states have changed in the time between the measured and actual timestamp.

\subsection{Estimation error in sensor fusion}
The goal of the sensor fusion algorithm is to estimate some states, internal or external, based on the sensor samples $\bm{y}(t_k)$ and control inputs $\bm{u}(t_k)$:
\begin{align}
    \hat{\bm x}(t_k) = \bm{g}(\bm{\hat{x}}(t_{k-1}); \bm{y}(t_{k}), \bm{u}(t_k)),
    \label{eq:sensor_fusion}
\end{align}
This is also known as non-linear state estimation.
The scope of this paper is limited to sensor fusion algorithms that estimate the position of objects, e.g. direct georeferencing, depth estimation, inertial navigation systems, etc.

In general, the error of the estimation of a system state variable $\bm z$ is defined as follows: 
\begin{align}
    \bm \epsilon &:= \bm z - \hat{\bm z}.
\end{align}

By assuming that the sensor fusion $\bm g$ is a linear combination of the measurements $\bm y$ can this estimation error can be represented by three components:
\begin{equation}
    \bm \epsilon = \bm \delta_\mathrm{sync}(\mu_\mathrm{sync}) + \bm \delta_\mathrm{sensor} + \bm \delta_\mathrm{systematic},
\end{equation}
where $\bm \delta_\mathrm{sync}$ is the estimation error caused by the synchronization error, here defined as the \textit{sync-induced error}. 
The variable $\bm \delta_\mathrm{sensor}$ is the estimation error caused by sensor noise, not to be confused with the sensor noise itself $\bm \epsilon_{sensor}$. The estimation error is the sensor noise passed through the sensor fusion algorithm, defined as the \textit{sensor-induced error}. The last source of error is $\bm \delta_\mathrm{systematic}$, which represent the systematic errors due to calibration, mounting, misalignment and so on. For the rest of this paper, such systematic errors are disregarded.

The Syncline model simplifies the sync-induced error to be a linear combination of the speed and the angular velocity of the robot scaled by the distance to the measured objects
\begin{align}
\bm \delta_\mathrm{sync}(\mu_\mathrm{sync}) &\approx \|{\bm v}_{eb}^e\|_2 \mu_\mathrm{sync} + d \cdot \|\bm \omega_{nb}^b\|_2\mu_\mathrm{sync},
    \label{eq:sync_error}
\end{align}
where $\|{\bm v}_{eb}^{e}\|_{2}$ is linear velocity of the robot, $d$ is the distance from the robot to the object of interest and $\|\bm \omega_{nb}^b\|_2$ is the angular velocity of the robot.
The left term of \cref{eq:sync_error} is an error caused by the linear distance between the robots actual position when a measurement was taken and its position at the time of the timestamp.
The right term is related to the difference in orientation of the robot between the time of the measurement sample and the time of the timestamp.
This difference in orientation, i.e. angle, is multiplied by the distance to the object of interest.
For small angles this is an approximation of the georeferencing error induced by the orientation error.

It is clear that the sync-induced error varies during the operation of the mobile robot, e.g. if there is no relative movement between the robot and the object of interest, then there is also no sync-induced error. 
An important characteristics of the sync-induced error is its \textit{expected worst-case} magnitude during a robot operation.
We define the expected worst-case sync-induced error as the error observed with the \textit{mean} of the synchronization error $\mu_\mathrm{sync}$ and the expected maximum linear and angular velocity. We also refer to this as the \textit{upper bound} and indicate it with an asterisk as follows:
\begin{equation}
    \delta_\mathrm{sync}^*(\tau) := v_\mathrm{max} \tau+ d \cdot\omega_\mathrm{max}\tau,
    \label{eq:sync_error_ub}
\end{equation}

Where $\tau$ is the worst-case synchronization error and, $v_\mathrm{max}$ and  $\omega_\mathrm{max}$ are norms of the expected maximum linear and angular velocities for the robot platform, respectively.

We define $\delta_\mathrm{sensor}^*$ as the expected worst-case of the sensor-induced error as follows
\begin{equation}
    \delta_\mathrm{sensor}^* := \sigma_p + \sigma_r + (\sigma_\Theta + \sigma_u)\cdot d,
    \label{eq:sensor_error_ub}
\end{equation}
where $\sigma_x$ is the standard deviation of the estimated variable $x$ $p$, $r$, $\Theta$ and $u$ represent localization, ranging, attitude and bearing sensors, respectively. 
In this context "expected" refers to that we use the standard deviation as the magnitude of the sensor error.
"Worst-case" refers to the fact that we are summing the magnitudes. This implies that the error vectors are positive scalar multiples of each other. Clearly the worst-case scenario is when all the error vectors are in the same direction.

\begin{figure}[tb]
  \centering
  \includegraphics[width=0.9\linewidth]{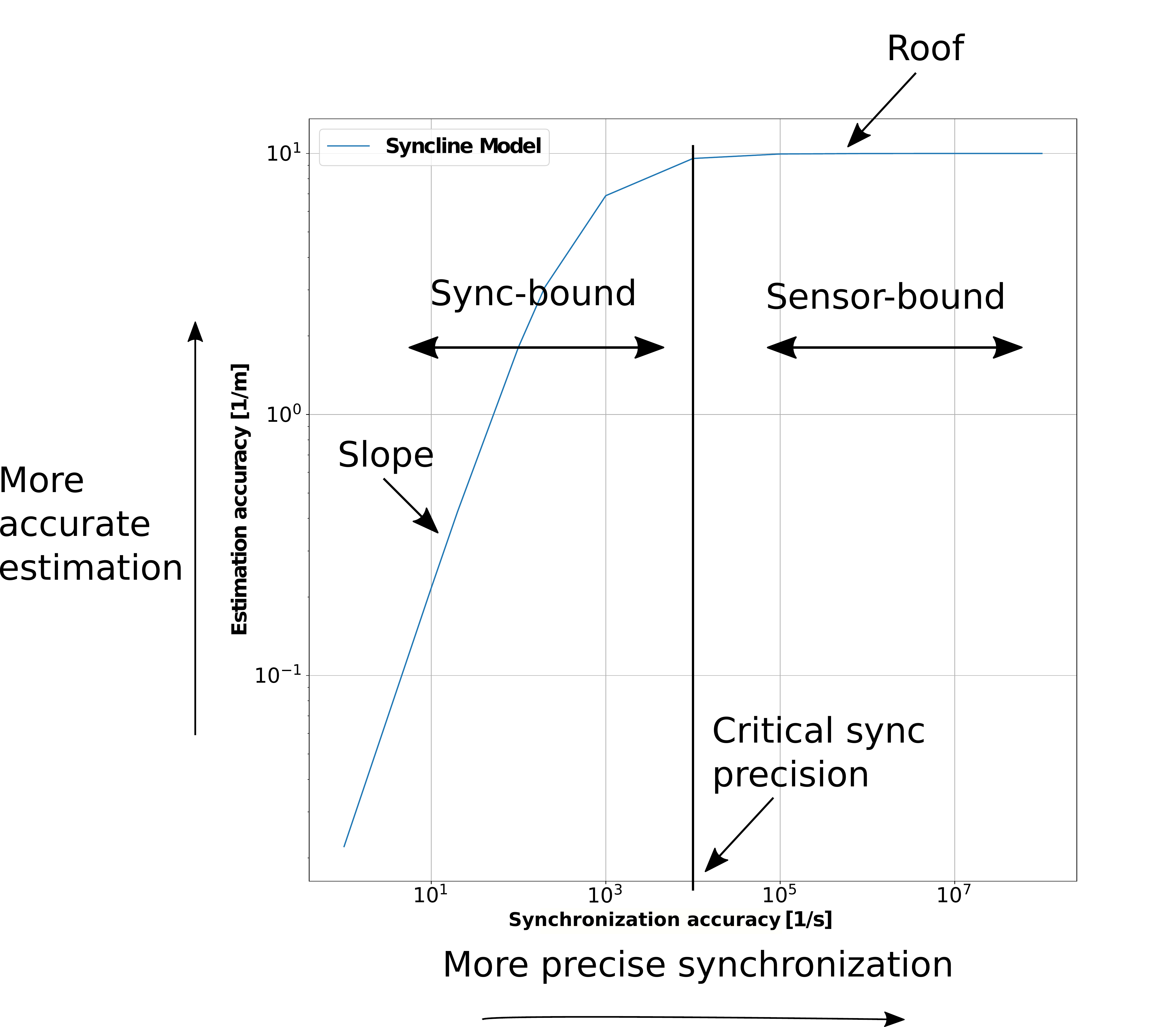}
  \vspace*{-0.3cm}
  \caption{\label{fig:sync-line}A Syncline plot}
  \vspace*{-0.1cm}
\end{figure}

Finally, we define the Syncline as
\begin{equation}
    \delta_\mathrm{syncline}(\tau) := \delta_\mathrm{sync}^*(\tau) + \delta_\mathrm{sensor}^*.
    \label{eq:syncline}
\end{equation}
The Syncline is an approximation of the worst case estimation accuracy as a function of the synchronization error.
\Cref{fig:sync-line} shows a plot of \cref{eq:syncline}, i.e. a Syncline.
Here, $\delta_\mathrm{sync}^*$ is calculated with the $v_\mathrm{max}$, $\omega_\mathrm{max}$ and $r$ estimated for a car, see Tab.~\ref{tab:examples}.
Also, $\delta_\mathrm{sensor}^*$ is set to the arbitrary value of \SI{0.1}{\meter}. 
This signifies that the sensors employed by the robot is capable of estimating positions with worst case error of \SI{0.1}{\meter}.
The vertical axis shows the \textbf{Estimation accuracy} which is the inverse of the error.
This is why the accuracy asymptotically reaches $10^1$ \SI{}{m^{-1}} which is the inverse of \SI{0.1}{\meter}.
The horizontal axis shows the \textbf{Synchronization Accuracy} which is the inverse the synchronization error 1/$\tau$, i.e. if the synchronization error is 1$\mu$s, then the synchronization accuracy 1/${\tau}$ is $10^6\mathrm{s}^{-1}$. We also define $\tau_\mathrm{crit}$ as the critical synchronization error which satisfies
\begin{equation}
    \delta_\mathrm{sync}^*(\tau_\mathrm{crit}) = \delta_\mathrm{sensor}^*. 
    \label{eq:t_crit}
\end{equation}
\Cref{fig:sync-line} indicates two regions along the horizontal axis.
The \textbf{Sync-bound} region is the region of high synchronization error where $\delta_\mathrm{sync}^* > \delta_\mathrm{sensor}^*$ which corresponds to $\tau > \tau_\mathrm{crit}$.
It is called sync-bound because the performance is mainly limited by the synchronization accuracy.
The \textbf{Sensor-bound} region is characterized by $\delta_\mathrm{sync}^* < \delta_\mathrm{sensor}^*$ and $\tau < \tau_\mathrm{crit}$. In this region, there is little use in improving the synchronization mechanism as intrinsic sensor noise is the limiting factor.
The border between the two regions indicates the critical synchronization error $\tau_{crit}$.
This is where $\delta_\mathrm{sync}^* = \delta_\mathrm{sensor}^*$ and $\tau = \tau_\mathrm{crit}$. 
In general, the system designer should achieve synchronization accuracy greater than the critical synchronization accuracy. 
This ensures that one gets maximum value from the sensors.

\Cref{fig:sync-line} also indicates the \textbf{slope} and the \textbf{roof} of the curve.
The slope shows how accuracy is increasing as we increase the synchronization accuracy.
The roof is the maximum obtainable accuracy for the complete sensor fusion system.

\section{Examples}
In this section, the Syncline model is evaluated on two examples of sensor fusion in mobile robots.
First, a state-space model of the sensor fusion algorithm is derived, it will serve as the "ground truth" and be the basis for the simulations of the system.
Then the Syncline model is derived and evaluated against the simulator.
Evaluating the model on real, experimental data, is left as future work.

To develop the Syncline model, we consider a set of robot platforms and their dynamics. Table \ref{tab:examples} shows the mobile robots and parameters used in these examples. The variable $d$ is the range and signifies the typical distance between the robot and the objects to be localized. The parameter $b$ stands for baseline which indicates the size of the robot and the maximum lever arms between the different sensors. The Syncline model is also dependent on the intrinsic sensor noise. We consider a set of commonly used sensors introduced in \Cref{tab:sensor_accuracy} where $\sigma(x)$ is the standard deviation of the measured state $x$. I.e. the intrinsic sensor noise.
\begin{table}[tb]
    \centering
    \vspace*{-0.2cm}
    \caption{\label{tab:examples}Robot platform parameters}
    \vspace*{-0.2cm}
    \begin{tabular}{ ccccc }
         \toprule
         \multicolumn{5}{c}{Robot dynamics} \\
         \midrule
         & $v_\mathrm{max}$~[$\frac{\mathrm{m}}{\mathrm{s}}$]& $\omega_\mathrm{max}$~[$\frac{\mathrm{deg}}{\mathrm{s}}$] & $d~[\SI{}{\meter}]$ & $b~[\SI{}{\meter}]$\\
         \midrule
          UAV Fixed Wing    & 21    & 77.9      & 100   & 1     \\
          UAV Multi Rotor   & 5     & 311.8     & 5     & 0.5   \\
          USV               & 5     & 17        & 30    & 5     \\
          AUV               & 30    & 8.7       & 5     & 5     \\
          Car               & 30    & 17.3      & 50    & 3     \\
          Large SV          & 2.5   & 4.5       & 1000  & 50    \\
          Small SV          & 4     & 17        & 1000  & 10    \\
         \bottomrule
    \end{tabular}
    \vspace*{-0.2cm}
\end{table}
\subsection{Example 1: Georeferencing with GNSS, INS and LiDAR}
In the first example we will consider is a georeferencing application where a mobile robot uses three sensors to estimate the global position of an external object. A GNSS receiver is used to measure its own global position, INS for attitude measurements and a LiDAR to measure range and bearing to the external object of interest.

\subsubsection{Analytical model}
In this example, we use a state space with 9 states comprising of the robots position and attitude as well as the position of the external object that we wish to georeference:
\begin{align}
    \bm{x} &= [\bm{p}_{eb}^e, \bm{\Theta}_{nb}, \bm{p}_{eo}^e]^T.
\end{align}

There are three sensors, each timestamps the measurements on its own clock i.e. $t_\mathrm{GNSS}$, $t_\mathrm{INS}$ and $t_\mathrm{LiDAR}$. The synchronization error $\mu_\mathrm{sync}$ is defined as relative the LiDAR timestamp. Thus, the GNSS synchronization error is defined as $\mu_\mathrm{sync}^\mathrm{gnss} = t_\mathrm{GNSS}-t_\mathrm{LiDAR}$.

The GNSS measurement is modeled as
\begin{align}
  \bm{y}_\mathrm{GNSS}(t) &= \hat{\bm p}_{eb_{g}}^e \\
  \hat{\bm p}_{eb_g}^e &= 
  %
  \bm p_{eb_g}^e + \bm{v}_{eb_g}^e\mu_\mathrm{sync}^\mathrm{gnss} + \bm \epsilon_p^\mathrm{gnss}.
  \label{eq:meas_gnss_base}
\end{align}
The first term is the true position of the GNSS sensor. The second term is the sync-induced error which is how much the GNSS sensor moves in the timestamping error period. The final term is the intrinsic sensor noise. The sync-induced error is a combination of the velocity of the GNSS receiver relative the ECEF frame and the synchronization error. The velocity can be expressed as a combination of the linear and angular velocity of the robot as follows:
\begin{align}
    \begin{split}
        \bm{v}_{eb_g}^e &= \frac{d}{dt}\bm{p}_{eb_g}^e = \frac{d}{dt}\left(\bm{p}_{eb_r}^e + \bm{p}_{b_{r}b_{g}}^e\right) \\ 
        &= \bm{v}_{eb_r}^e + \dot{\bm{R}}_{eb_r}\bm{p}_{b_{r}b_{g}}^{b_r}.
    \end{split}
    \label{eq:meas_gnss_frames}
\end{align}

The rotation from the ECEF frame to the robot frame is composed of the two rotations $\bm{R}_{eb_r} = \bm{R}_{en} \bm{R}_{nb_r}$. We assume that $\bm{R}_{en}$ is constant during the time synchronization error $\mu_{sync}^{gnss}$. We can then express the time derivative of that rotation as: $\dot{\bm{R}}_{eb_r} = \bm{R}_{eb_r}\bm{S}(\bm{\omega}_{nb_r}^{b_r})$. Inserted into \cref{eq:meas_gnss_frames} yields 
\begin{align}
    \begin{split}
        \bm{v}_{eb_g}^e = \bm{R}_{eb_r} \left(\bm{v}_{eb_r}^{b_r} + \bm{S}(\bm{\omega}_{eb_r}^{b_r})\bm{p}_{b_{r}b_{g}}^{b_r} \right).
    \end{split}
\end{align}
further inserting into \cref{eq:meas_gnss_base} gives us
\begin{align}
  \begin{split}
  \hat{\bm p}_{eb_g}^e &= \bm p_{eb_g}^e 
  + \bm R_{eb_r}\left(\bm v_{eb_r}^{b_r} + \bm S(\bm \omega_{nb_r}^{b_r})\bm p_{b_rb_g}^{b_r}\right)\mu_\mathrm{sync}^\mathrm{ins} 
  + \bm \epsilon_{p}^\mathrm{gnss} \\
  \end{split}
  \label{eq:meas_gnss}
\end{align}
The INS measures the attitude of the robot as follows
\begin{align}
  \bm{y}_\mathrm{INS}(t) &= \hat{\bm R}_{nb_r} \\
  \hat{\bm R}_{nb_r} &= \bm{R}_{nb_r} + \bm{R}_{nb_r}\bm{S}(\bm{\omega}_{nb_r}^{b_r})\mu_\mathrm{sync}^\mathrm{ins} + \bm \epsilon_{\Theta}^\mathrm{ins}.
  \label{eq:meas_ins}
\end{align}
The first term is the true attitude. The second term is the sync-induced error, it is the time derivative of the attitude multiplied with the timestamp offset. This assumes that the angular velocity is constant during this time period. The last term is the intrinsic sensor noise.
\begin{align}
  \bm{y}_{\mathrm{L}}(t) &= \hat{\bm p}_{b_lb_o}^{b_r} \\
  \hat{\bm p}_{b_lb_o}^{b_r} &= \bm R_{b_rb_l}(\bm p_{b_lb_o}^{b_r} + \bm \epsilon_p^\mathrm{lidar})
    \label{eq:meas_lidar}
\end{align}
I.e the LiDAR outputs a position vector from itself $b_l$ to the object of interest $b_o$. 
In reality the LiDAR outputs a 3D point cloud with azimuth, elevation and range for each point. 
This signal must pass through a signal processing algorithm which performs object detection.
We assume the output of this object detection is a range $r$ and a bearing vector $\bm{n}^{b_l}_{b_lb_o}$ defined as 
\begin{equation}
  \bm{n}^{b_l}_{b_lb_o} := 
  \begin{bmatrix}
    \cos(\Psi)\cos(\alpha) &
    \sin(\Psi)\cos(\alpha) & 
    -\sin(\alpha) 
  \end{bmatrix}^\intercal
\end{equation}
where $\Psi$ is azimuth and $\alpha$ is elevation such that the total vector from $\{b_l\}$ to $\{b_o\}$ can be stated
\begin{align}
  \bm p_{b_lb_o}^{b_l} = r \cdot \bm{n}^{b_l}_{b_lb_o} = \bm R_{b_lb_o}(\alpha, \Psi) \cdot \bm d_{b_lb_o}
\label{eq:bear_posvec}
\end{align}
and where 
\begin{equation}
    \bm R_{b_lb_o}(\alpha, \Psi) = 
        \begin{bsmallmatrix}
        \cos(\Psi)\cos(\alpha)  & -\sin(\Psi)   &  \sin(\alpha) \cos(\Psi) \\
        \sin(\Psi)\cos(\alpha)  & \cos(\Psi)    &  \sin(\alpha) \sin(\Psi) \\
        -\sin(\alpha)           & 0             &  \cos(\alpha)
        \end{bsmallmatrix}
        \label{eq:R_bearing}
\end{equation}
and
\begin{equation}
    \bm d_{b_lb_o} = 
    \begin{bmatrix}
        r & 0 & 0
    \end{bmatrix}^\intercal.
\end{equation}
Notice that the LiDAR measurement has no sync-induced error. This is because we have defined synchronization error as relative to the LiDAR timestamp.

The sensor fusion is defined by the function $\bm g$ introduced in  \cref{eq:sensor_fusion}. It combines INS, GNSS and LiDAR measurements to estimate the global ECEF position of the object, which we write as
\begin{equation}
     \bm g(\bm{x}(t_{k-1}); \bm{y}(t_{k})) = \hat{\bm{p}}_{eo}^e(t_k).
\end{equation}
This is a series of rotations and translations:
\begin{align}
    \hat{\bm{p}}_{eo}^e(t_k) = \hat{\bm{p}}_{eb_g}^e + \bm{R}_{en}\hat{\bm{R}}_{nb_r}(\bm{p}_{b_gb_l}^{b_r} + \hat{\bm{p}}_{b_lb_o}^{b_r}).
    \label{eq:analytic_direct_georef}
\end{align}
This can again be expressed in terms of the true quantities, sync-induced errors and sensor noise by substituting in \cref{eq:meas_gnss}, \cref{eq:meas_ins} and \cref{eq:meas_lidar}.

\subsubsection{Syncline model}
For the sync-induced error, we use the $v_{\mathrm{max}}$ and ${\omega_\mathrm{max}}$ from \Cref{tab:examples}.
For the sensor error, we use the $\sigma$-values from \Cref{tab:sensor_accuracy}.
 
\begin{table}[tb]
    \centering
    \caption{\label{tab:sensor_accuracy}Sensor noise}
    \vspace*{-0.1cm}
    \begin{tabular}{ccccc}
        \toprule
        \multicolumn{5}{c}{GNSS} \\
        \midrule
        & uBlox & uBlox & Trimble & Trimble \\
        & F9P PVT & F9P RTK & R12 DGNSS & R12 RTK \\
        \midrule
        $\sigma(p)$  & \SI{1.5}{\metre} & \SI{1}{\centi\metre} & \SI{0.25}{\metre} & \SI{8}{\milli\metre}   \\
        \toprule
        \multicolumn{5}{c}{INS} \\
        \midrule
        & SBG & SBG & Kongsberg &\\
        & Ellipse & Apogee & MRU5 &  \\
        \midrule
        $\sigma(\phi)$ & \SI{0.1}{\degree} &   \SI{0.008}{\degree} &    \SI{0.002}{\degree} & \\
        $\sigma(\theta)$  & \SI{0.1}{\degree} &  \SI{0.008}{\degree} &\SI{0.002}{\degree} &\\
        $\sigma(\psi)$ & \SI{0.2}{\degree}  &  \SI{0.03}{\degree} & \SI{0.002}{\degree}  & \\
        \toprule
        \multicolumn{5}{c}{LiDAR} \\
        \midrule
        & Velodyne & Velodyne & RIEGL &  Faro\\
        & Alpha Prime & HDL32E & VUX1-UAV & Focus Plus \\
        \midrule
        $\sigma( r)$ & \SI{4}{\centi\metre}& \SI{2}{\centi\metre}  & \SI{1}{\centi\metre} & \SI{1}{\milli\metre} \\
        $\sigma(\Psi)$ & \SI{0.1}{\degree} & \SI{0.08}{\degree}& \SI{0.006}{\degree} & \SI{0.005}{\degree}\\
        $\sigma(\alpha)$ & \SI{0.2}{\degree} & \SI{0.08}{\degree}&\SI{0.006}{\degree} & \SI{0.005}{\degree}\\
        \toprule
        \multicolumn{5}{c}{USBL and MBE} \\
        \midrule
        & Kongsberg & Sensodyne & Kongsberg &  Sonic\\
        & HIPAP502 & USBL7000 & M3Sonar & 2026 MBE \\
        \midrule
        $\sigma( r)$ & \SI{2}{\centi\metre}& \SI{1.5}{\centi\metre}  & \SI{1}{\centi\metre} & \SI{1}{\milli\metre} \\
        $\sigma(\Psi)$ & \SI{0.06}{\degree} & \SI{0.04}{\degree}& \SI{0.9}{\degree} & \SI{0.45}{\degree}\\
        $\sigma(\alpha)$ & \SI{0.06}{\degree} & \SI{0.04}{\degree}&\SI{0.5}{\degree} & \SI{0.45}{\degree}\\
        \bottomrule
    \end{tabular}
\end{table}%
\begin{figure}[tb]
    \centering
    \vspace*{-0.2cm}
    \includegraphics[width=0.8\linewidth]{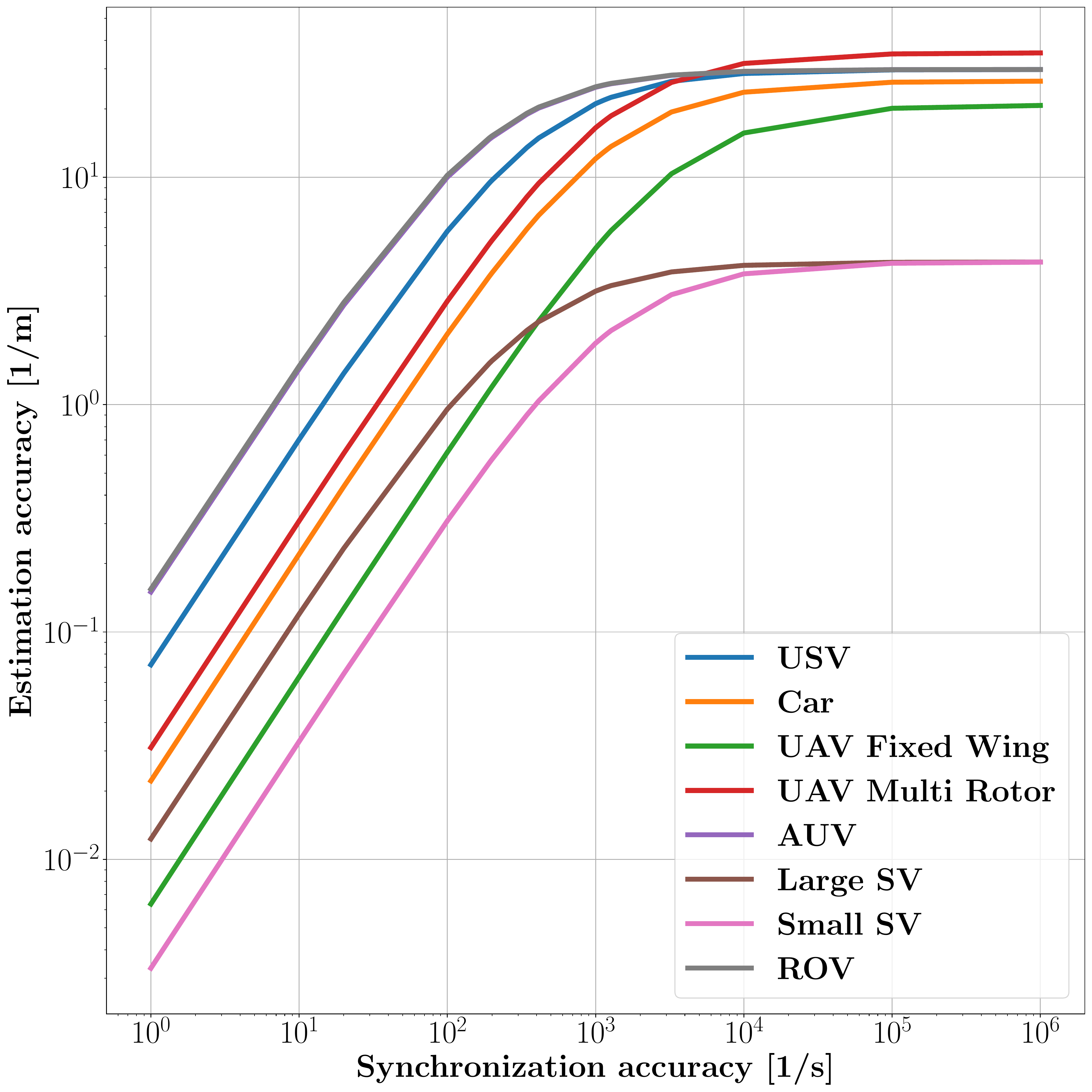}
    \vspace*{-0.2cm}
    \caption{\label{fig:syncline-direct-georef}Syncline for direct georeferencing with uBlox F9P RTK, MRU5 and VUX1-UAV for various robot platforms}
    \vspace{-0.2cm}
\end{figure}%
\Cref{fig:syncline-direct-georef} compares the Synclines for different mobile robot platforms when using uBlox F9P RTK, MRU5 and VUX1-UAV.
There are several interesting things to note.
Firstly, the Synclines have different roofs.
The roof is the accuracy achieved when you are only limited by the intrinsic sensor accuracy.
The reason why the robots have different roofs is that they are assumed to be georeferencing objects at different distances. Secondly, the length of the slope of the Synclines varies. 
This is related to the sensitivity to synchronization errors. 
The shorter the roof (i.e. a lower $\tau_{\mathrm{crit}}$), the more sensitive a system is to synchronization errors. This depends on the robot dynamics and the intrinsic sensor noise. 

\begin{figure}[tb]
    \centering
    \includegraphics[width=0.8\linewidth]{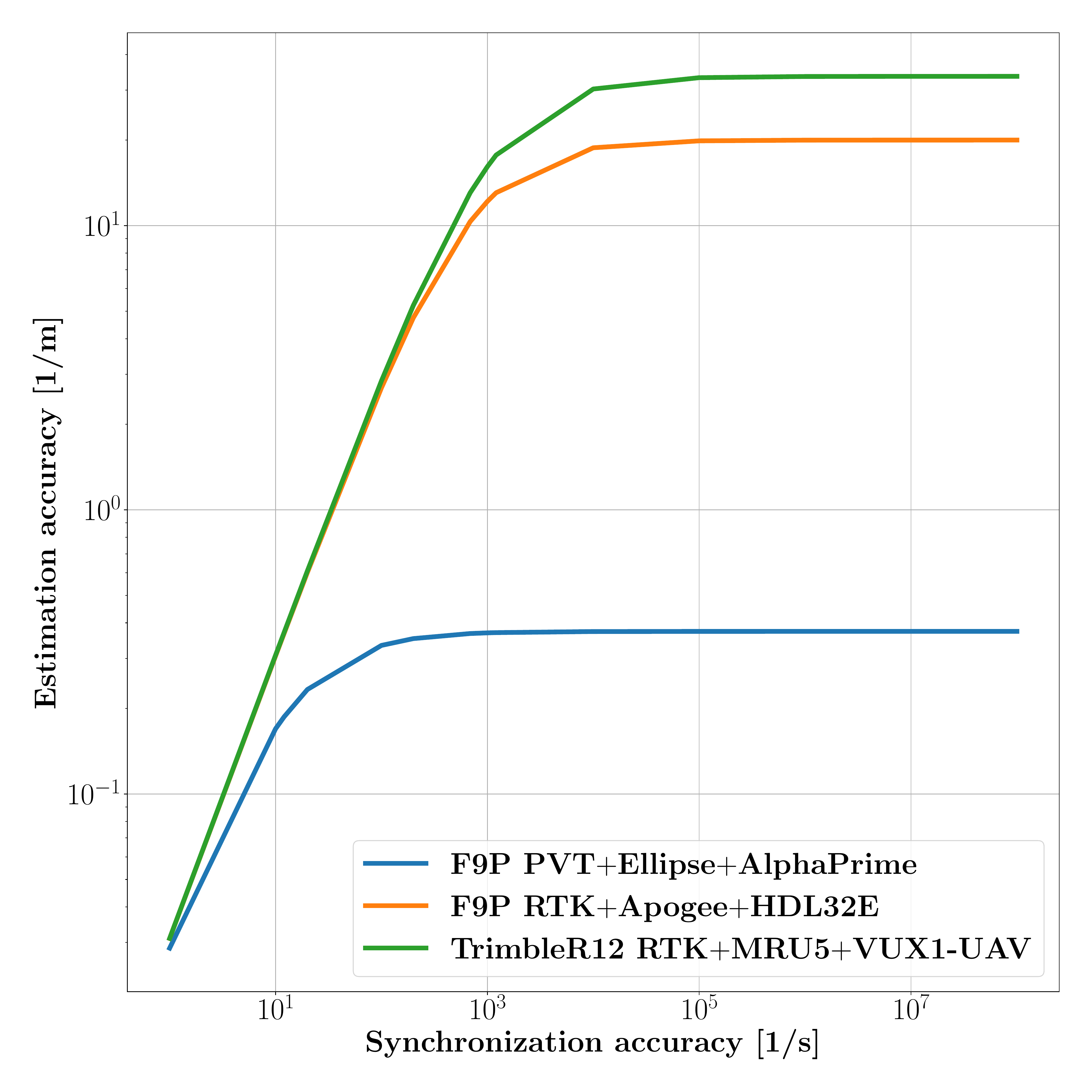}
    \vspace*{-0.2cm}
    \caption{\label{fig:syncline-sensors-direct-georef}Syncline model for UAV Fixed Wing with different sensor payloads}
    \vspace*{-0.2cm}
\end{figure}

\Cref{tab:georef_tau} reports the $\tau_\mathrm{crit}$ for the different sensors individually.
The synchronization error requirements for the sensor fusion is dictated by the sensor with the lowest $\tau_\mathrm{crit}$.
\begin{table}[tb]
    \centering
    \caption{\label{tab:georef_tau} Minimum synchronization error $\tau_\mathrm{crit}$}
    \vspace*{-0.2cm}
    \begin{tabular}{ ccccc }
         \toprule
         \multicolumn{5}{c}{Robot platform} \\
         \midrule
         Sensor & USV & Car & Multi Rotor& Fixed Wing\\
         \midrule
         F9P PVT & 520ms & 86ms & 520ms & 123ms \\
         F9P RTK & 3ms & 577us & 3ms & 824us \\
         Ellipse & 9ms & 4ms & 663us & 2ms  \\
         MRU5 & 130us & 67us & 9us & 38us\\
         AlphaPrime  & 11ms  & 5ms & 2ms & 3ms  \\
         Vux1    & 1ms & 385us & 333us & 157us  \\
         \bottomrule
    \end{tabular}
    \vspace*{-0.2cm}
\end{table}%
In \Cref{fig:syncline-sensors-direct-georef}, a plot with a few different combinations of the sensor payload for a UAV platform is presented.
Intuitively, changing the sensor payload raises or lowers the roofs of the Synclines, which depends on the intrinsic sensor noise only. 
It also changes the length of the roofs which is related to $\tau_{crit}$. 
The shorter the roof, the more sensitive the system is to synchronization error.

\begin{figure}[tb]
    \centering
    \includegraphics[width=0.8\linewidth]{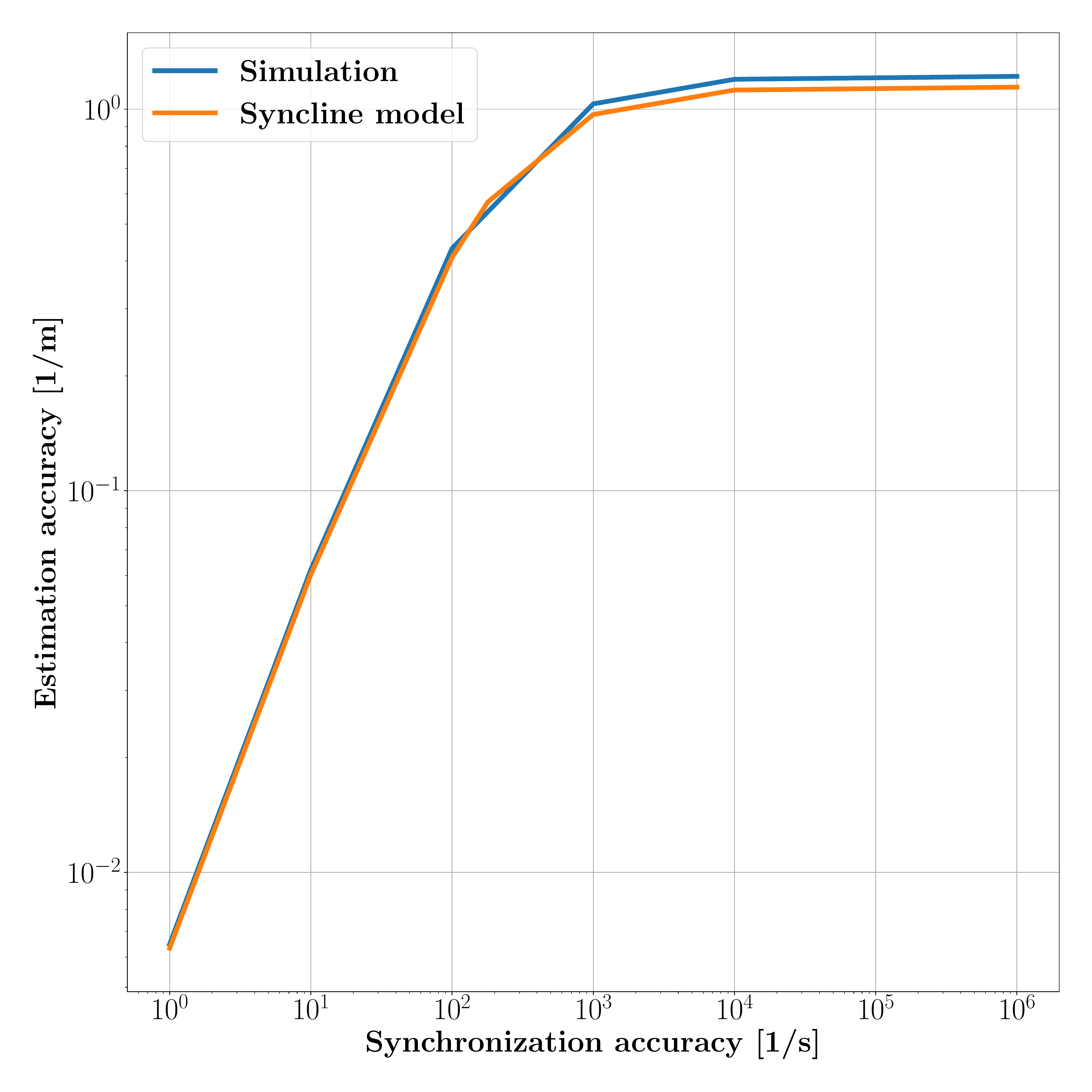}
    \vspace*{-0.2cm}
    \caption{\label{fig:syncline-sim-compare-direct-georef}Syncline model for UAV Fixed Wing compared with simulations}
    \vspace*{-0.2cm}
\end{figure}

To verify the accuracy of the Syncline model, we have made a simulator based on the analytical formulation in \cref{eq:analytic_direct_georef} is also implemented.
We simulate a fixed wing UAV which maneuvers according to \Cref{tab:examples} trying to georeference a fixed object using senors with parameters in \Cref{tab:sensor_accuracy}. 
For each synchronization error, the worst case georeferencing error is found and compared with the Syncline in \Cref{fig:syncline-sim-compare-direct-georef}. 
Clearly, the Syncline model is adept for modeling direct georeferencing.

\subsection{Example 2: Underwater survey with USBL, MBE, INS and GNSS}
In the second example, we consider a subsea survey system consisting of a surface vessel (SV) and an autonomous underwater Vehicle (AUV). The surface vessel has a GNSS, INS$_\mathrm{sv}$, and Ultra-short baseline acoustic positioning system (USBL) receiver.
The AUV has a USBL transponder, a INS$_\mathrm{auv}$ and a Multibeam Echosounder (MBE).
This system is inspired by the authors in~\cite{ffi}
The goal is to estimate global position of the AUVs MBE footprint $\bm{p}_{eo}^{e}$. This is achieved by performing direct georeferencing twice.
First, the global position of the AUV $\bm{p}_{eb_\mathrm{auv}}^{e}$ is estimated based on the GNSS measurements $\bm{\hat{p}}_{eb_\mathrm{sv}}^{e}$, the INS$_\mathrm{sv}$ measurements $\bm{\hat{R}}_{nb_\mathrm{sv}}$ and the USBL measurements $\bm{\hat{p}}_{b_\mathrm{sv}b_\mathrm{auv}}$, all taken at the surface vessel. 
This position estimate is fused with the INS$_\mathrm{auv}$ measurement $\bm{\hat{R}}_{nb_\mathrm{auv}}$ and the range-and-bearing measurements from the MBE to the ocean floor $\bm{\hat{p}}_{b_\mathrm{auv}o}^{b_\mathrm{auv}}$.

As before, both an analytical model and the Syncline model are derived. 
\subsubsection{Analytical model}
The state space consists of the position and attitude of the SV, the position and attitude of the AUV and the position of the MBE footprint.
\begin{align}
  \bm{X} = [\bm{p}_{eb_\mathrm{sv}}^{e}, \bm{\Theta}_{nb_\mathrm{sv}}, \bm{p}_{eb_\mathrm{auv}}^{e} \bm{\Theta}_{nb_\mathrm{auv}}, \bm{p}_{eo}^{e}]^\intercal. 
\end{align}
The system has five sensors which are all timestamped with their respective clock.
The synchronization error is defined as relative the MBE measurements.
The GNSS, INS$_\mathrm{sv}$ and USBL measurements of the SV are combined into a single virtual position sensor for the AUV based on direct georeferencing. 
Internal to the virtual sensor, the synchronization error is modeled as relative to the USBL measurement. 

The GNSS and INS measurements are derived in \cref{eq:meas_gnss} and \cref{eq:meas_ins}, respectively.
The USBL measurements is a vector from  the receiver on the SV to the transponder on the AUV, as follows:
\begin{align}
    \bm{y}_\mathrm{USBL} &= \hat{\bm p}_{b_{urx}b_\mathrm{utp}}^{b_\mathrm{sv}} \\
    \hat{\bm p}_{b_{urx}b_\mathrm{utp}}^{b_\mathrm{sv}} &= \bm p_{b_{urx}b_\mathrm{utp}}^{b_\mathrm{sv}} + \bm \epsilon_p^\mathrm{usbl}.
      \label{eq:meas_usbl}
\end{align}

The virtual AUV position measurement is defined as follows:
\begin{align}
    \bm{y}_\mathrm{AUVPOS} &= \hat{\bm p}_{eb_\mathrm{utp}}^e \\
    \begin{split}
    \hat{\bm p}_{eb_\mathrm{utp}}^e &= \hat{\bm{p}}_{eb_g}^e 
    +\bm{R}_{en}\hat{\bm{R}}_{nb_\mathrm{sv}}(\bm{p}_{b_gb_{urx}}^{b_\mathrm{sv}} + \hat{\bm{p}}_{b_{urx}b_\mathrm{utp}}^{b_\mathrm{sv}}) \\
    &+  \bm R_{eb_{auv}}\left(\bm v_{eb_{auv}}^{b_{auv}} + \bm S(\bm \omega_{nb_{auv}}^{b_{auv}})\bm p_{b_{utp}b_{auv}}^{b_{auv}}\right)\mu_\mathrm{sync}^\mathrm{auvpos}.
    \end{split}
    \label{eq:meas_auv_pos}
\end{align}

The first term is the measured position of the GNSS receiver at the SV. The second term is the measured orientation of the SV combined with the mounting lever arms for the sensors on the SV. The last term is the sync-induced error caused by the movement of the AUV during the timestamping offset between the virtual AUV position measurement and the AUV MBE measurement. 

The INS$_{auv}$ is modeled as follows
\begin{align}
  \bm{y}_\mathrm{INS_\mathrm{auv}}(t) &= \hat{\bm R}_{nb_\mathrm{auv}} \\
  \hat{\bm R}_{nb_\mathrm{auv}} &= \bm{R}_{nb_\mathrm{auv}} + \bm{R}_{nb_\mathrm{auv}}\bm{S}(\bm{\omega}_{nb_r}^{b_r})\mu_\mathrm{sync}^\mathrm{ins_{sv}} + \bm \epsilon_{\Theta}^\mathrm{ins_{sv}}
  \label{eq:meas_ins_uv}
\end{align}
This is similar to \cref{eq:meas_ins}. The error in the attitude measurement is a combination of the sync-induced error and the intrinsic sensor noise. 

The MBE measurement is analogous to the LiDAR measurement in  \cref{eq:meas_lidar} in Example 1, and outputs the position vector $\hat{\bm p}_{b_{mbe}b_o}^{b_\mathrm{auv}}$ as a range and a bearing as described in \cref{eq:bear_posvec}. The global position of the footprint can now be expresses in terms of the virtual AUV position measurement, the INS$_\mathrm{auv}$ measurement and the MBE measurement
\begin{align}
\begin{split}
    \hat{\bm{p}}_{eo}^e(t_k) &= \hat{\bm{p}}_{eb_g}^e + \bm{R}_{en}\hat{\bm{R}}_{nb_\mathrm{sv}}(\bm{p}_{b_gb_{urx}}^{b_\mathrm{sv}} + \hat{\bm{p}}_{b_{urx}b_\mathrm{utp}}^{b_\mathrm{sv}})\\
    &+ \bm{R}_{en}\hat{\bm{R}}_{nb_\mathrm{auv}}(\bm{p}_{b_\mathrm{utp}b_l}^{b_\mathrm{auv}} + \hat{\bm{p}}_{b_{mbe}b_o}^{b_\mathrm{auv}}).
\end{split}
\label{eq:analytic_subsea}
\end{align}
\subsubsection{Syncline model}
The Syncline model is easily extended to account for the dynamics of both the SV and the AUV as follows
\begin{equation}
    \delta_\mathrm{sync}^*(\tau) = (v_\mathrm{max}^{\mathrm{sv}}+ v_\mathrm{max}^{\mathrm{auv}}+ d^{\mathrm{sv}} \omega_\mathrm{max}^{\mathrm{sv}}+d^{\mathrm{auv}} \omega_\mathrm{max}^{\mathrm{auv}})\cdot\tau.
    \label{eq:sync_error_survey}
\end{equation}
This is the sum of the two Synclines, one for the direct georeferencing of the AUV from the SV, and one for the direct georeferencing of the seabed from the AUV.

The sensor-induced errors are expressed as follows:
\begin{align}
\begin{split}
    \delta_\mathrm{sensor}^* &= \sigma_{p_\mathrm{sv}} + \sigma_{r_\mathrm{sv}} 
    + (\sigma_{\Theta_\mathrm{sv}} + \sigma_{u_\mathrm{sv})}\cdot d_\mathrm{sv} \\
    &+ \sigma_{p_\mathrm{auv}} + \sigma r_\mathrm{auv}
    + (\sigma_{\Theta_\mathrm{auv}} + \sigma_{u_\mathrm{auv}}) \cdot d_\mathrm{auv}.
\end{split}
\end{align}
\begin{figure}[tb]
    \centering
    \vspace*{-0.2cm}
    \includegraphics[width=0.8\linewidth]{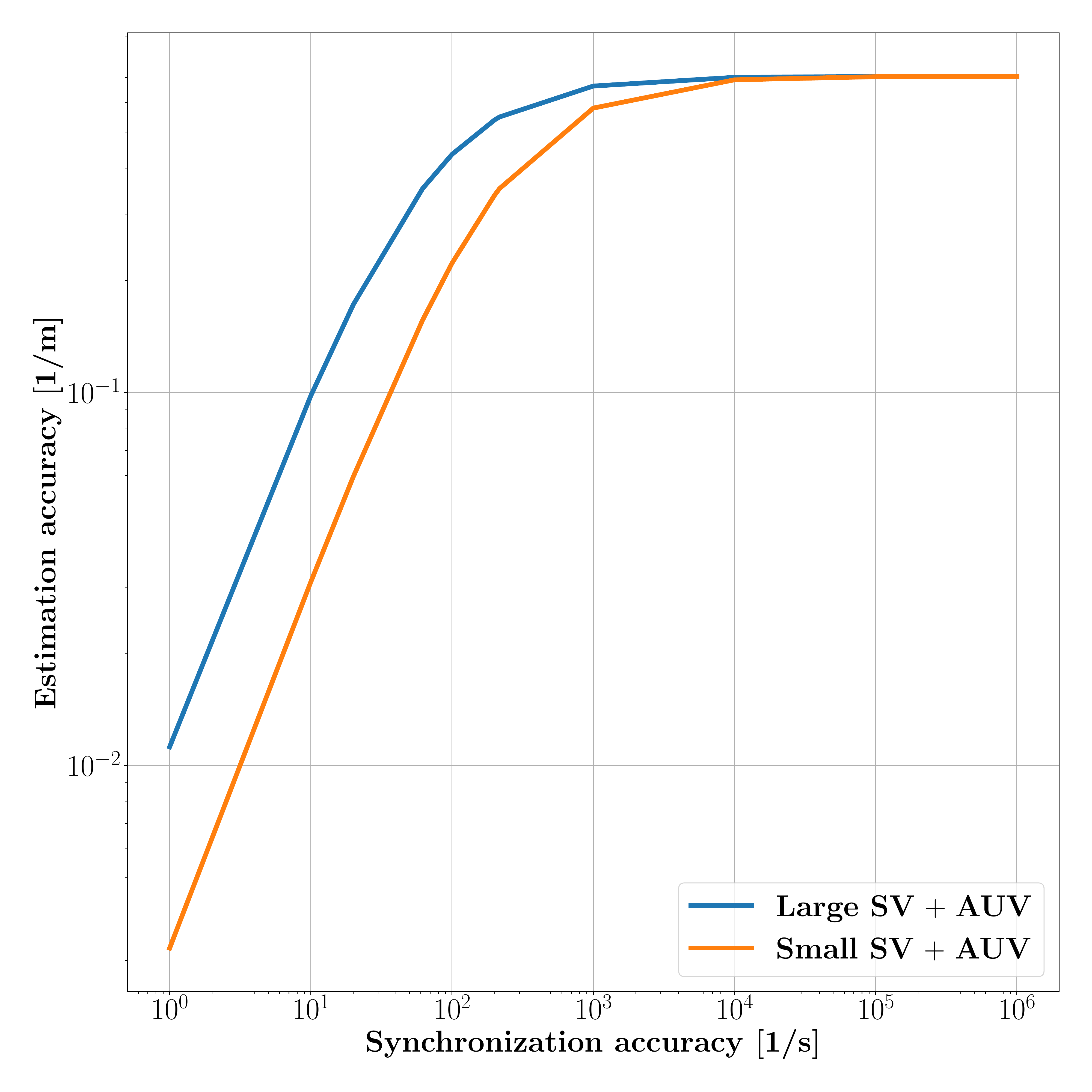}
    \vspace*{-0.2cm}
    \caption{\label{fig:syncline-survey}Synclines for underwater survey with SV and AUV using Trimble R12 RTK GNSS, MRU5 INS, Sensodyne Gyro USBL7000 and Sonic 2026 MBE}
    \vspace*{-0.2cm}
\end{figure}%
The Syncline model is plotted in \Cref{fig:syncline-survey} for both a large SV and a small SV performing subsea survey together with an AUV.
The AUV is assumed to be at a 1000m depth and 30m above the sea floor.
Both systems has roofs at around 1.67m.
This is expected since they are equipped the same sensor payload and are at the same distance from the object of interest.
The system with large SV has a longer roof which is due to the fact that it has slower dynamics and thus less sensitivity to synchronization errors. For the large SV $\tau_{crit}=16$ ms while for the small SV $\tau_{crit}=4$ ms. 

\begin{figure}[tb]
    \vspace*{-0.2cm}
    \centering
    \includegraphics[width=0.8\linewidth]{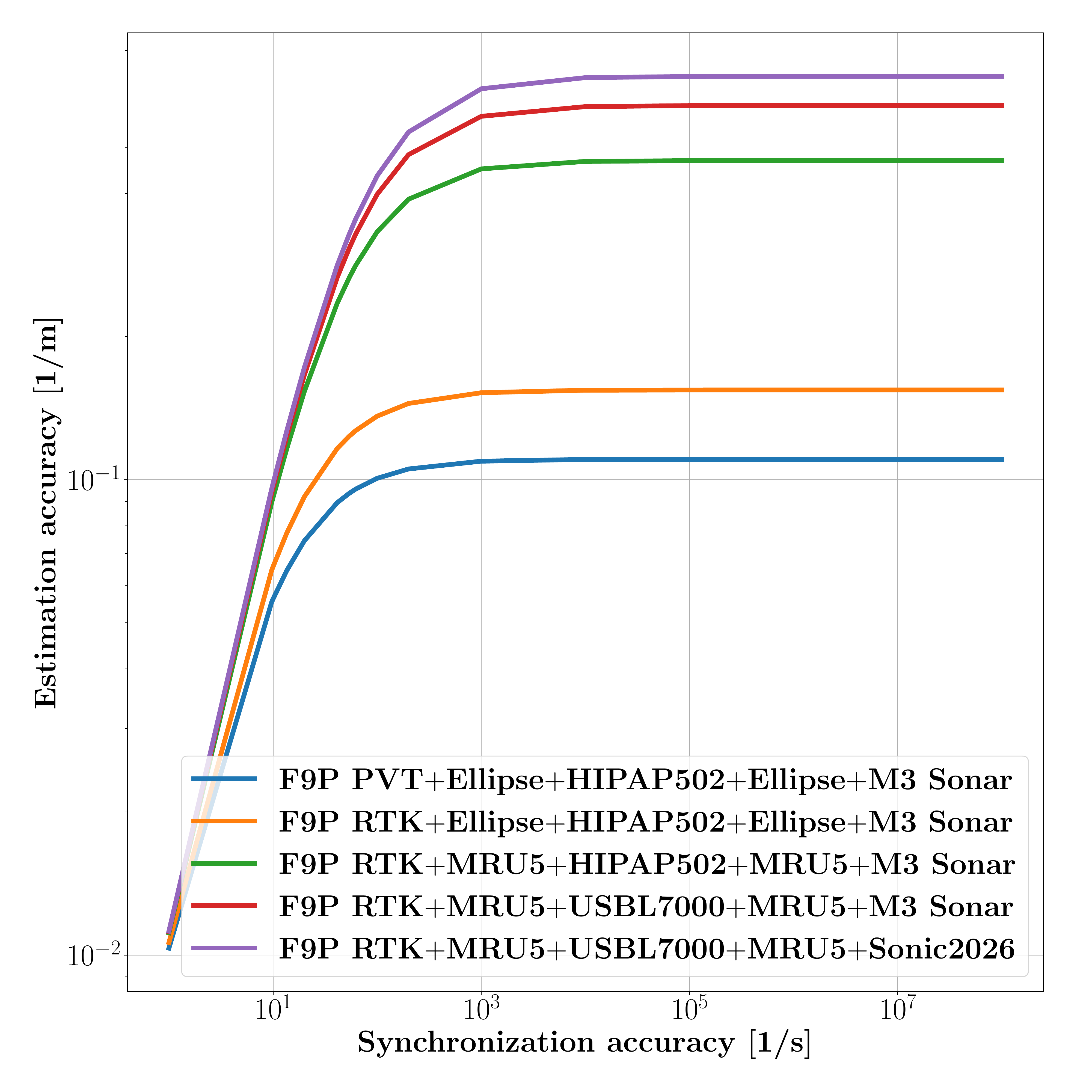}
    \vspace*{-0.2cm}
    \caption{\label{fig:syncline-sensors-survey}Syncline model for Large SV and AUV with different sensor payloads}
    \vspace*{-0.2cm}
\end{figure}

In \Cref{fig:syncline-sensors-survey}, the Synclines for a system with a large SV and a AUV are compared for four different sensor payloads. 
Clearly, replacing the GNSS receiver with a better one does comparatively little with the performance compared to replacing the INS or USBL.
\Cref{tab:survey_tau} reports the critical synchronization per sensor. This illustrates sensitivity of each sensor to synchronization error. For instance, the uBlox F9P PVT which only has around \SI{1.5}{\meter} accuracy is not sensitive to synchronization error on such slow moving robot platforms as SVs and AUVs. The MRU5 which can estimate the robot attitude with an accuracy of \SI{0.002}{\degree} is much more sensitive, even on these slow platforms. 

\begin{figure}[tb]
    \centering
    \vspace*{-0.2cm}
    \includegraphics[width=0.8\linewidth]{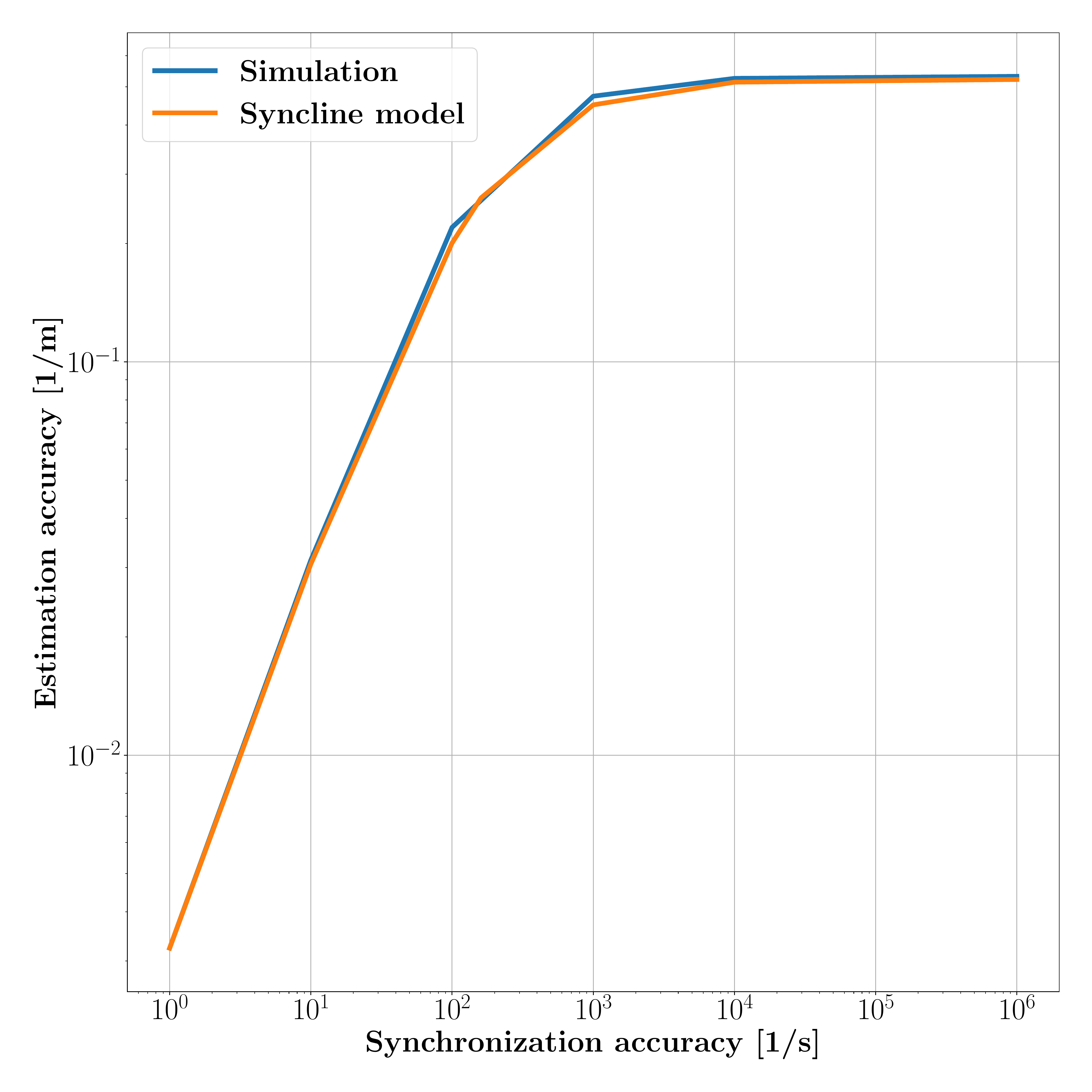}
    \vspace*{-0.2cm}
    \caption{\label{fig:syncline-sim-compare-survey}Evaluation of Syncline model for Small SV and AUV against a simulator. THe sensors used are uBlox F9P RTK, MRU5, HIPAP520 and Sonic 2026}
    \vspace*{-0.2cm}
\end{figure}

\begin{table}[tb]
    \centering
    \vspace*{-0.2cm}
    \caption{\label{tab:survey_tau} Critical synchronization error}
    \vspace*{-0.2cm}
    \begin{tabular}{ cccc }
        \toprule
        \multicolumn{4}{c}{Robot platform} \\
        \midrule
        $\tau_\mathrm{crit}$ & Large SV & Small SV & AUV \\
        \midrule
        uBlox F9P PVT & 1039.230ms & 649.519ms & - \\
        uBlox F9P RTK & 6.928ms & 4.330ms & - \\
        SBG Ellipse & 52.754ms & 14.217ms & 19.335ms \\
        MRU5 & 0.746ms & 0.201ms & 0.273ms \\
        HIPAP502 & 18.521ms & 4.991ms & 9.713ms \\
        USBL7000 & 12.368ms & 3.333ms & 6.727ms \\
        Sonic2026 & 137.071ms & 36.940ms & 50.385ms \\
        M3 Sonar & 221.857ms & 59.790ms & 82.776ms \\
         \bottomrule
    \end{tabular}
    \vspace*{-0.2cm}
\end{table}%
In \Cref{fig:syncline-sim-compare-survey} the Syncline model is evaluated against a simulator based on the analytical model derived in \cref{eq:analytic_subsea}. Again, there is no significant deviation between the Syncline model and the analytical simulation.
%


\section{Conclusions}
The Syncline model is a simple visual model which allows system designers to explore the interaction between errors introduced by lacking sensor synchronization and errors introduced by intrinsic sensor noise.
Synclines are derived for a number of different robot platforms and sensor payloads.
The model is then evaluated against a mobile robot simulator for two different sensor fusion algorithms.
The results are promising showing that the Syncline model is accurate for direct georeferencing and mapping applications.

\vspace{6pt} 




\bibliography{references}
\bibliographystyle{IEEEtran}

\end{document}